\documentclass[aps,prb,twocolumn,showpacs,nofootinbib]{revtex4}
\usepackage{graphicx}
\usepackage{bm}
\usepackage{amssymb}
\usepackage{epstopdf}
\usepackage{amsmath}
\usepackage{amstext}
\usepackage{amsthm}
\usepackage{amsfonts}
\usepackage{latexsym}
\usepackage{hyperref}
\usepackage[margin,index]{fixme}
\usepackage[T1]{fontenc}

\renewcommand{\emph}[1]{{\it #1}}

\newcommand{\eps}{\varepsilon}

\begin{document}

\title{Dynamical symmetry breaking with optimal control: reducing the number of pieces}
\author{Matthew J. M. Power}
\affiliation{Centre for Theoretical Atomic, Molecular and Optical Physics,
Queen's University Belfast, Belfast BT7 1NN, United Kingdom}
\author{Gabriele De Chiara}
\affiliation{Centre for Theoretical Atomic, Molecular and Optical Physics,
Queen's University Belfast, Belfast BT7 1NN, United Kingdom}

\begin{abstract}
We analyse the production of defects during the dynamical crossing of a mean-field phase transition with a real order parameter. When the parameter that brings the system across the critical point changes in time according to a power-law schedule, we recover the predictions dictated by the well-known Kibble--Zurek theory.
For a fixed duration of the evolution, we show that the average number of defects can be drastically reduced for a very large but finite system, by optimising the time dependence of the driving using optimal control techniques. Furthermore, the optimised protocol is robust against small fluctuations.
\end{abstract}
\maketitle
\section{Introduction}
Non-equilibrium dynamics of many-body systems have been the subject of intensive investigation in statistical physics.
While for systems in quasi-static equilibrium, fluctuation--dissipation relations can be applied, these in general do not hold for systems driven out of equilibrium \cite{Calabrese2005}. Interest in the dynamics of many-body systems has recently focused on the thermalisation of isolated systems~\cite{Polko_review} and on the evolution of  systems that are brought to the verge of a critical point or that are made to cross it. 

The divergence of the reaction times of the system in the critical region causes every attempt to drive the system adiabatically to be useless. As a consequence,  the lack of sufficient time for the system to adapt to the rapid changes of its temperature or some other parameter gives rise to the creation of topological defects. These can be kinks, domain walls or even more complicated structures depending on the dimensionality of the system \cite{mermin}.

It was Kibble~\cite{Kibble} who first introduced this idea of quickly crossing a symmetry-breaking transition for explaining structure formation in the early universe. Later, Zurek~\cite{Zurek} proposed the same mechanism in a condensed matter setting in which theoretical predictions might be more easily assessed in experiments. This theory, now known as the {\it Kibble--Zurek} mechanism, predicts that the rate of production of defects is proportional to a power $\chi$
of the rate of change of the parameter in the system that drives it across the transition. 
The power $\chi$ is related to the critical exponents describing the scaling of physical quantities close to the critical point~\cite{Zurek}. 

These simple, yet powerful, causality arguments, leading to universal scaling relations, were later extended to quantum phase transitions at zero temperature~\cite{QPT}.
The observation of the Kibble--Zurek mechanism has been proposed and tested in many physical realisations: superfluid helium~\cite{Zurek1996,Hendry}, liquid crystals~\cite{Chuang1991}, arrays of Josephson junctions~\cite{Kavo2000,Dziarmaga2002}, superconducting films~\cite{Maniv2003}, ion Coulomb crystals~\cite{delCampo2010,DeChiara2010,exp_ions}, Bose-Einstein condensates~\cite{Uhlmann2007,Saito2007,delCampo2011,Sabbatini2011,Ferrari2013} and solid-state hexagonal manganite materials~\cite{Spaldin2012}. See Ref.~\cite{Adolfo2013} for a recent review.

Here we show that the number of defects produced during the crossing of a mean-field transition can be significantly reduced by applying simple optimal control techniques. 
First, we revisit the dynamics of a classical second order phase transition led by a control parameter for the $\varphi^4$ model in a one-dimensional lattice as in \cite{Laguna1998}.  Moreover the system is assumed to be in contact with a thermostat of very low temperature (see details below) so that the smearing of the transition is extremely small and the correlation length and relaxation time exhibit well defined maxima. It has been demonstrated that this model describes the dynamics of quasi one-dimensional ion crystals subject to laser cooling~\cite{DeChiara2010}.

For a constant-rate linear quench $\eps(t) \sim t/\tau_Q$ we recover the original Kibble-Zurek scaling in which the number of defects grows with the $1/4$ power of the rate $1/\tau_Q$. We then move to non-linear quenches $\eps(t)\sim (t/\tau_Q)^\alpha$ and show agreement with previous results~\cite{Mondal2009,Collura2010,Krapivsky,Chandran2012}. If we restrict the evolution to a fixed time $T$, we thus find an optimal power $\alpha$ such that the number of defects produced is minimised~\cite{Barankov2008,Barankov2013}.
We go beyond this scenario and apply an adaptation of the chopped random basis (CRAB) algorithm~\cite{CRAB} to optimise the functional dependence $\eps(t)$ in a fixed time $T$ with the goal of reducing the number of defects created. We find more than a $40\%$ decrease in the average number of defects created thus demonstrating the effectiveness of optimal control techniques in the open-system scenario. Moreover, as we show below, the number of defects created is robust against small perturbations in the time-dependence of $\eps(t)$. 
Our scheme has potential applications in the preparation of many-body systems in the equilibrium configuration of ordered phases with the aim of producing the largest domains.

The paper is organised as follows: in Sec.~\ref{sec:model} we discuss in detail the model we consider and its numerical simulation; in Sec.~\ref{sec:quenches} we present the results for both linear and non linear quenches and we recover the Kibble-Zurek scaling; in Sec.~\ref{sec:optimal} we explain our optimisation technique and show the results for the optimised quenches and the reduced average number of defects; finally, in Sec.~\ref{sec:conclusions} we summarise and conclude.

\section{The model}
\label{sec:model}

Following Ref.~\cite{Laguna1998}, we consider a one-dimensional mean field theory with real order parameter $\varphi(x,t)$ and real coordinate $x$ that depends on time $t$. Close to the critical point, the potential energy term is of the Landau's $\varphi^4$ form:
\begin{equation}
\label{eq:landau}
V(\varphi) = \frac 18\left[\varphi^4(x,t) - 2\eps(t)\varphi^2(x,t)\right],
\end{equation}
where $\eps(t)$ is a mass term or, equivalently, the reduced temperature; in a more general sense, it is the parameter of the system that drives the transition. 
When $\eps(t)\le 0$ the Landau potential has only one real minimum $\varphi=0$ that corresponds to the symmetric vacuum in the disordered phase. When $\eps(t)>0$ the potential $V(\varphi)$ is characterised by two symmetry--broken minima of the ordered phase: $\varphi = \pm \sqrt{\eps}$. The critical point thus corresponds to $\eps(t)=0$.
The order parameter dynamics in space and time fulfils the Ginzburg--Landau partial differential equation:
\begin{equation}
\label{eq:GL}
\left[\frac{\partial^2}{\partial t^2}+\eta\frac{\partial}{\partial t}-\frac{\partial^2}{\partial x^2}\right] \varphi(x,t)+\frac{\partial V(\varphi)}{\partial \varphi} = \vartheta(x,t)
\end{equation}
where $\eta$ and $\vartheta(x,t)$ are the phenomenological dissipation rate and Langevin force, respectively, that ensure thermalisation for constant $\eps(t)$.
In this paper we will consider dimensionless units such that $\eta=1$.
Model \eqref{eq:GL} has been employed by Laguna and Zurek to verify numerically Kibble--Zurek scaling, in the simplest possible scenario~\cite{Laguna1998}. When $\eps(t)$ is changed rapidly in time from a negative to a positive value, the order parameter exhibits spontaneous local decay towards either the positive or negative minimum of $V(\varphi)$. Crucially,  in spatially separated regions, the order parameter $\varphi(x,t)$ may develop an opposite sign giving rise to defects.  

We assume the Langevin forces to be random variables with no spatial or temporal correlations:
\begin{equation}
\langle\vartheta(x,t)\vartheta(x',t') \rangle=2\eta\theta\delta(x-x')\delta(t-t'),
\end{equation}
where $\theta$ is an effective temperature of the environment that is in contact at all times with the system.
In accordance with Laguna and Zurek\cite{Laguna1998} we choose $\theta=0.01$. This low temperature value ensures that the density of defects in the form of domain walls that might arise from thermal fluctuations for $\eps>0$ is negligible\cite{Krumhansl}. This means that practically all the defects that we count at the end of the process are formed during the fast quench of $\eps(t)$.
We also assume the system to be in the over-damped regime, corresponding to the parameter $\eta$ being larger than all the real eigenfrequencies of Eq.~\eqref{eq:GL}. Under this assumption, the order parameter will always monotonically decay to its steady state when $\eps$ ceases to change.

For our numerical simulations, we employ the finite--difference method and the velocity Verlet algorithm to simulate the dynamics of Eq.~\eqref{eq:GL}. 
As in \cite{Laguna1998}, we initially take $N=2^{14}$ spatial grid points with a periodic domain. This relatively large number of points allows us to recover in a clear and unambiguous way the Kibble--Zurek scaling. The initial condition is $\varphi(x,t_{in})=0$ where $t_{in}$ is the initial time.

During the quench protocol, $\eps(t)$ changes from $\eps(t_{in})=-2$ at the initial time $t_{in}$ to $\eps(t_{fin})=5$ at the final time $t_{fin}$ such that the total time is $T=t_{fin}-t_{in}$.
For this choice of the initial and final values of $\eps$, the average value of the equilibrium order parameter coincides with the minimum of the potential energy \eqref{eq:landau} with only small fluctuations.
 For each simulation, we count the number of defects $N_D$ as the number of zeros of the order parameter $\varphi(x,t_{fin})$ (counting the pairs of adjacent grid points where $\varphi$ changes sign). We average $N_D$ over no less than $N_{av}=10^3$ different realisations of the Langevin  forces. This is enough to obtain small statistical fluctuations in the average results.

\begin{figure}[t!]
\begin{center}
\includegraphics[width=0.8 \columnwidth ]{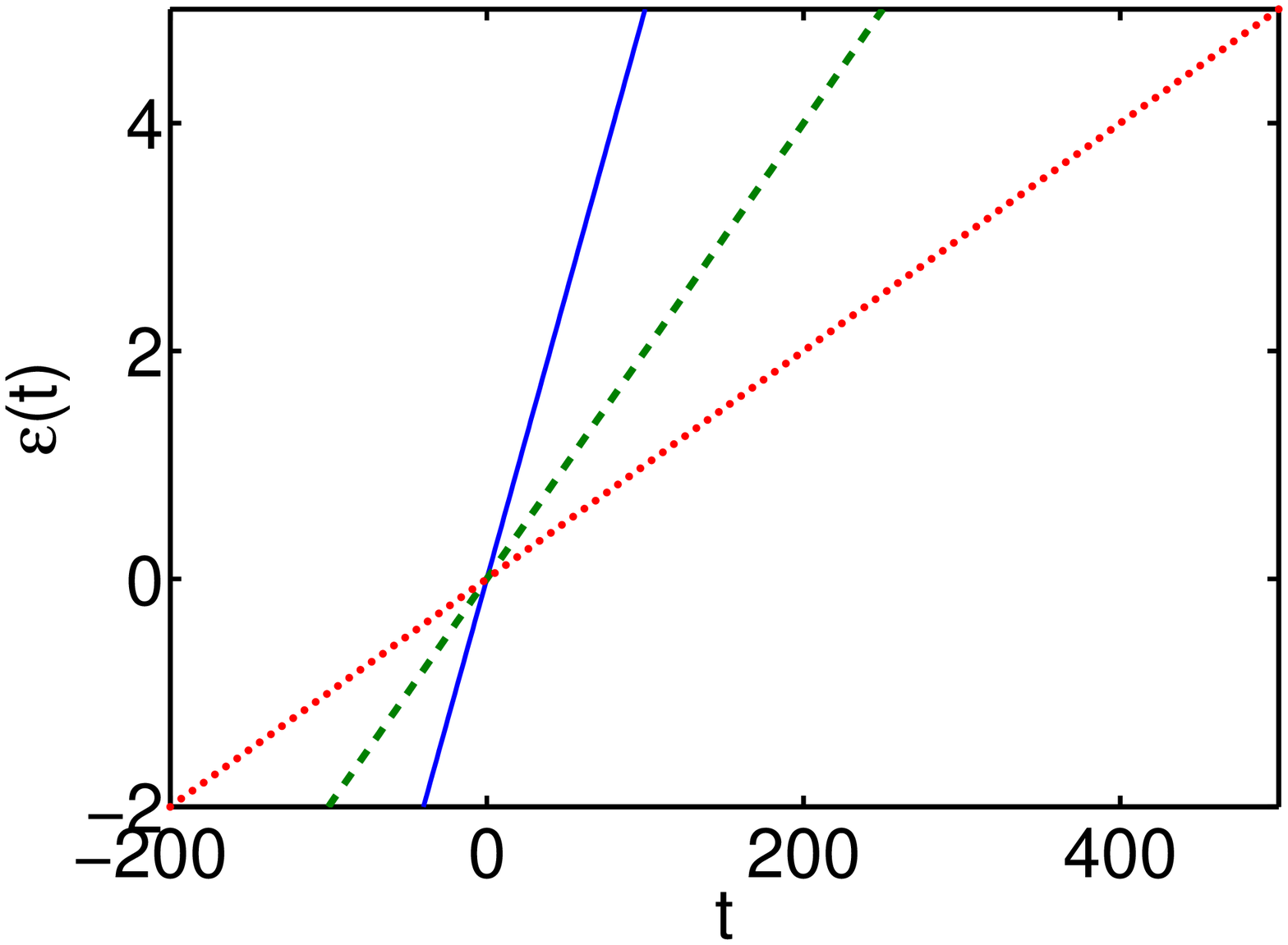}
\includegraphics[width=0.8 \columnwidth ]{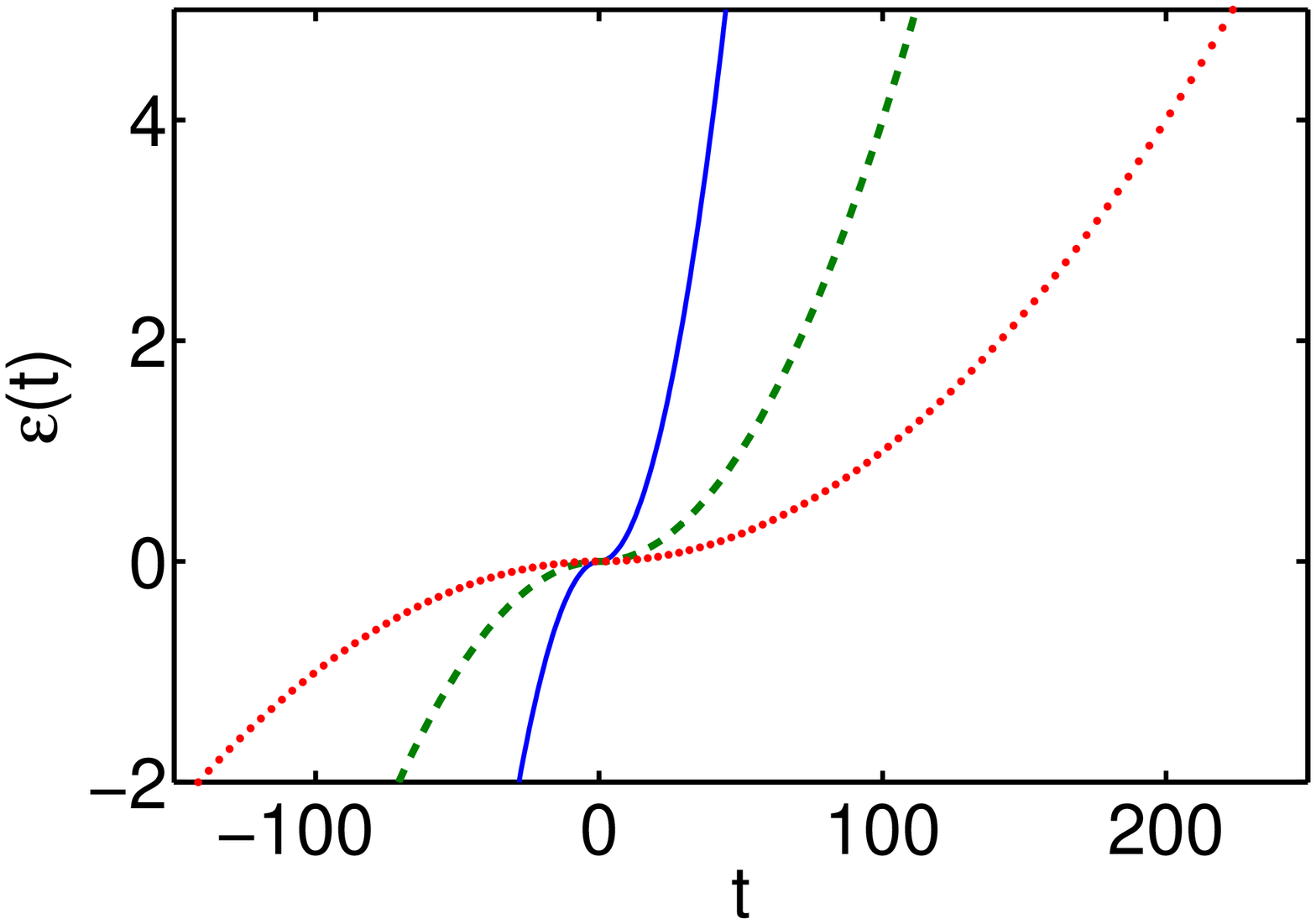}
\caption{(Color online) Time dependence of the function $\eps(t)$ from Eq.~\eqref{eq:quench1} for $\alpha=1$ (top) and $\alpha=2$ (bottom) for $\tau_Q=20$ (solid line), 50 (dashed line) and 100 (dotted line).}
\label{fig:eps1}
\end{center}
\end{figure}

\section{Linear and non linear quenches}
\label{sec:quenches}

We begin our investigation by testing our model and its finite-difference implementation for linear and non linear quenches of the form:
\begin{equation}
\label{eq:quench1}
\eps(t) = \frac{t}{|t|} \left(\frac{|t|}{\tau_Q} \right)^\alpha,
\end{equation}
where we defined the rate $\tau_Q^{-1}$ of crossing the critical point. Notice that the critical point $\eps=0$ is reached always at $t=0$. In order to ensure the correct initial and final values we set:
\begin{eqnarray}
\label{eq:tintfin}
t_{in}= -2^{1/\alpha} \tau_Q;
\quad\quad
t_{fin}= 5^{1/\alpha} \tau_Q
\end{eqnarray}
Therefore, with these settings, the total time $T=t_{fin}-t_{in}$ depends on both $\alpha$ and $\tau_Q$.
A few examples of the time dependence of $\eps(t)$ are shown in Fig.~\ref{fig:eps1}.

For the protocol in Eq.~\eqref{eq:quench1}, the predicted scaling for the average number of defects reads~\cite{Mondal2009,Collura2010,Chandran2012}:
\begin{eqnarray}
\label{eq:nd}
N_D &\sim& \left(\frac{1}{\tau_Q}\right)^\chi\\
\chi&=& \frac{\alpha\nu}{\alpha\mu+1} 
\label{eq:chi}
\end{eqnarray}
where, for the Ginzburg-Landau model we consider, $\nu=1/2$ and $\mu=1$ are the correlation length and relaxation time critical exponents of the mean-field universality class.
For $\alpha=1$, Eq.~\eqref{eq:chi} gives $\chi=1/4$ as first derived by Zurek~\cite{Zurek}.

\begin{figure}[t]
\begin{center}
\includegraphics[width=0.8 \columnwidth ]{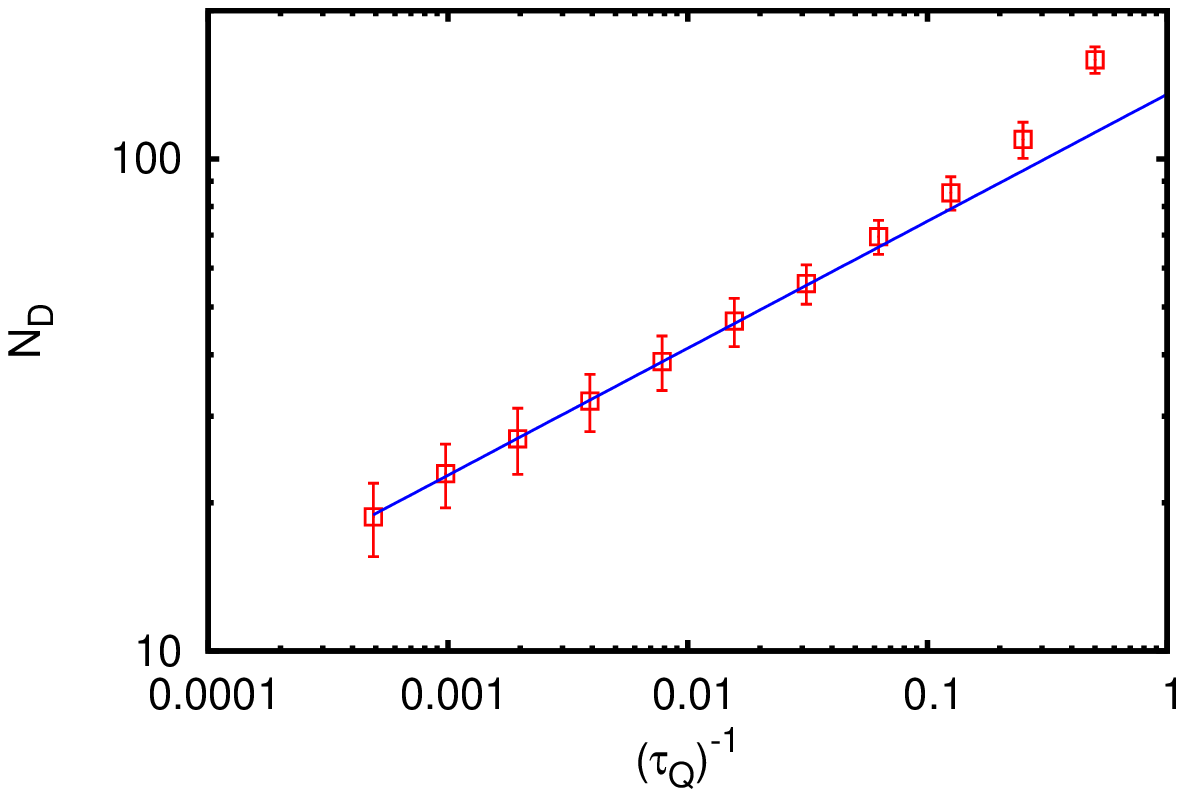}
\includegraphics[width=0.8 \columnwidth ]{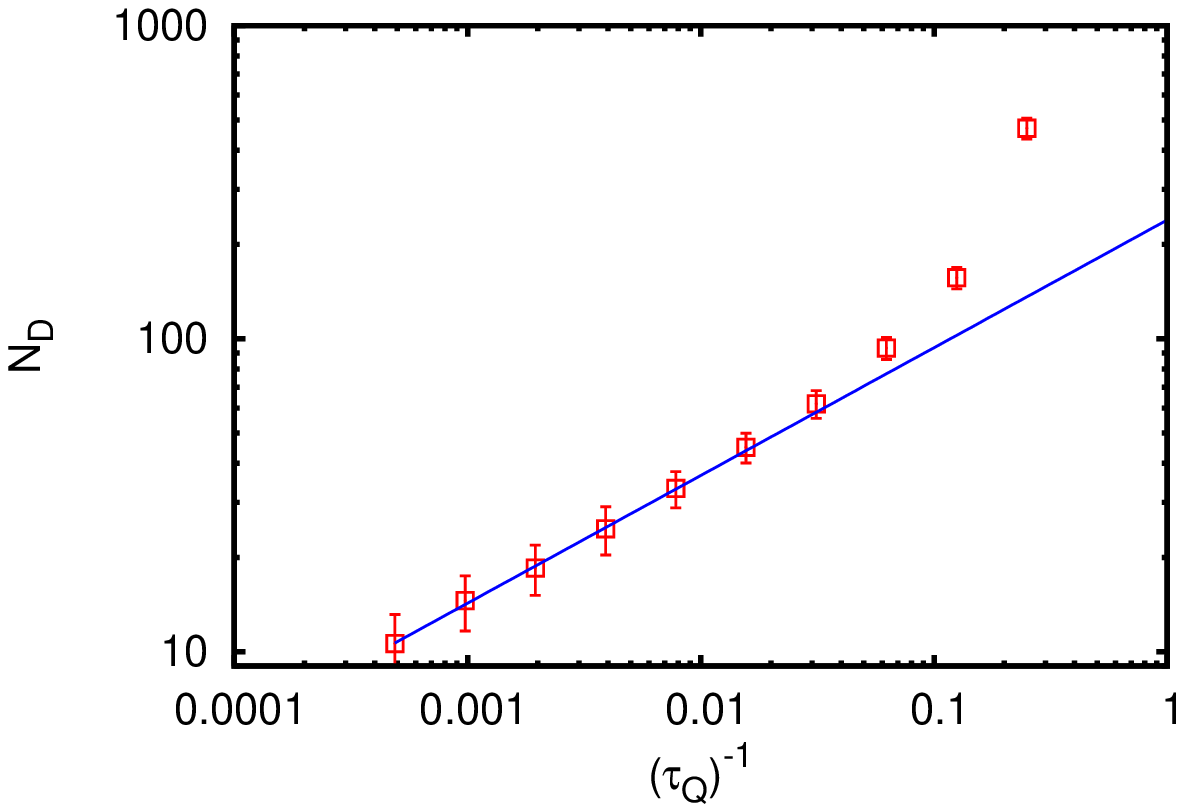}
\caption{(Color online) Scaling of the average number of defects $N_D$ (points) as a function of $1/\tau_Q$ in log-log scale for $\alpha=1$ (top) and $\alpha=4$ (bottom). Also shown are the best-fit lines according to prediction \eqref{eq:nd}.}
\label{fig:quench_lin_sq}
\end{center}
\end{figure}

After performing numerical simulations of Eq.~\eqref{eq:GL} with $\alpha=1$ and measuring the average number of defects $N_D$ we find the results shown in Fig.~\ref{fig:quench_lin_sq} as a function of the quench time $\tau_Q$. The scaling of $N_D$ with $\tau_Q^{-1}$ is linear over more than 2 order of magnitudes thanks to the large size of the integration domain. The best-fit result, in the linear region, for the scaling exponent is $\chi_{fit} =0.258\pm 0.004$ that is very close to the expected result $\chi=0.25$.

\begin{figure*}[htbp]
\begin{center}
(a)\hspace{6cm}(b)\\
\includegraphics[width=6.1cm]{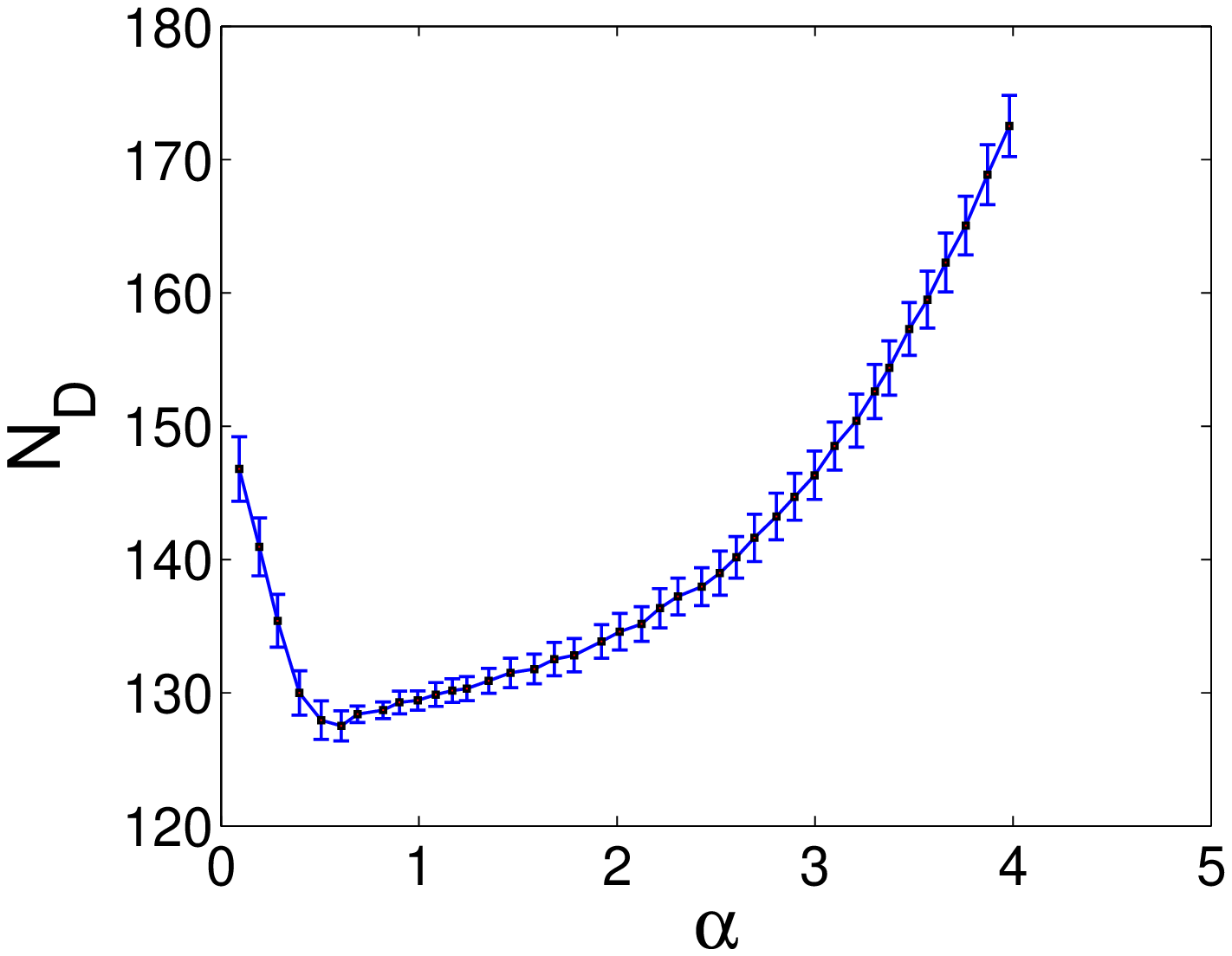}
\includegraphics[width=6.1cm]{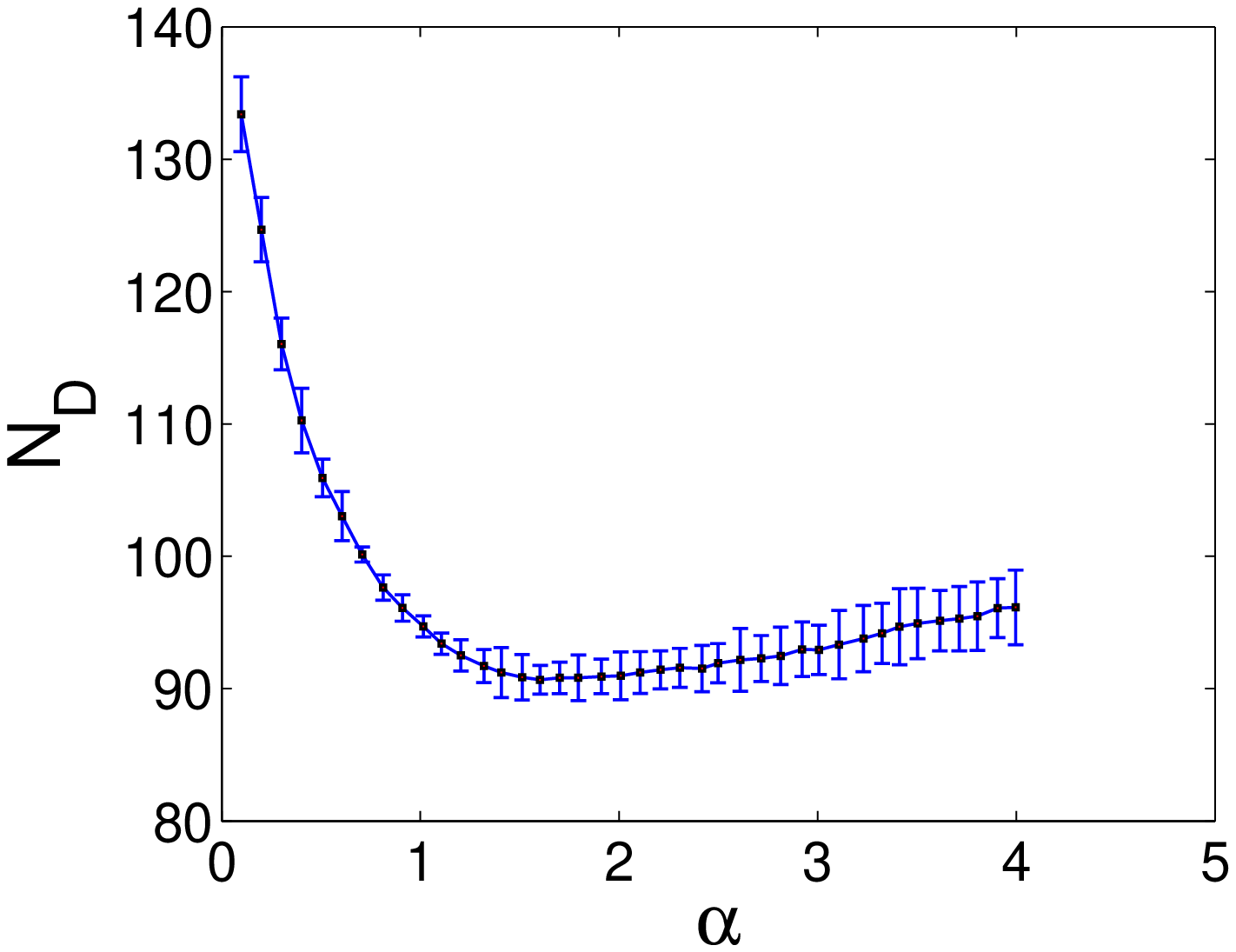}\\
(c)\hspace{6cm}(d)\\
\includegraphics[width=6.1cm]{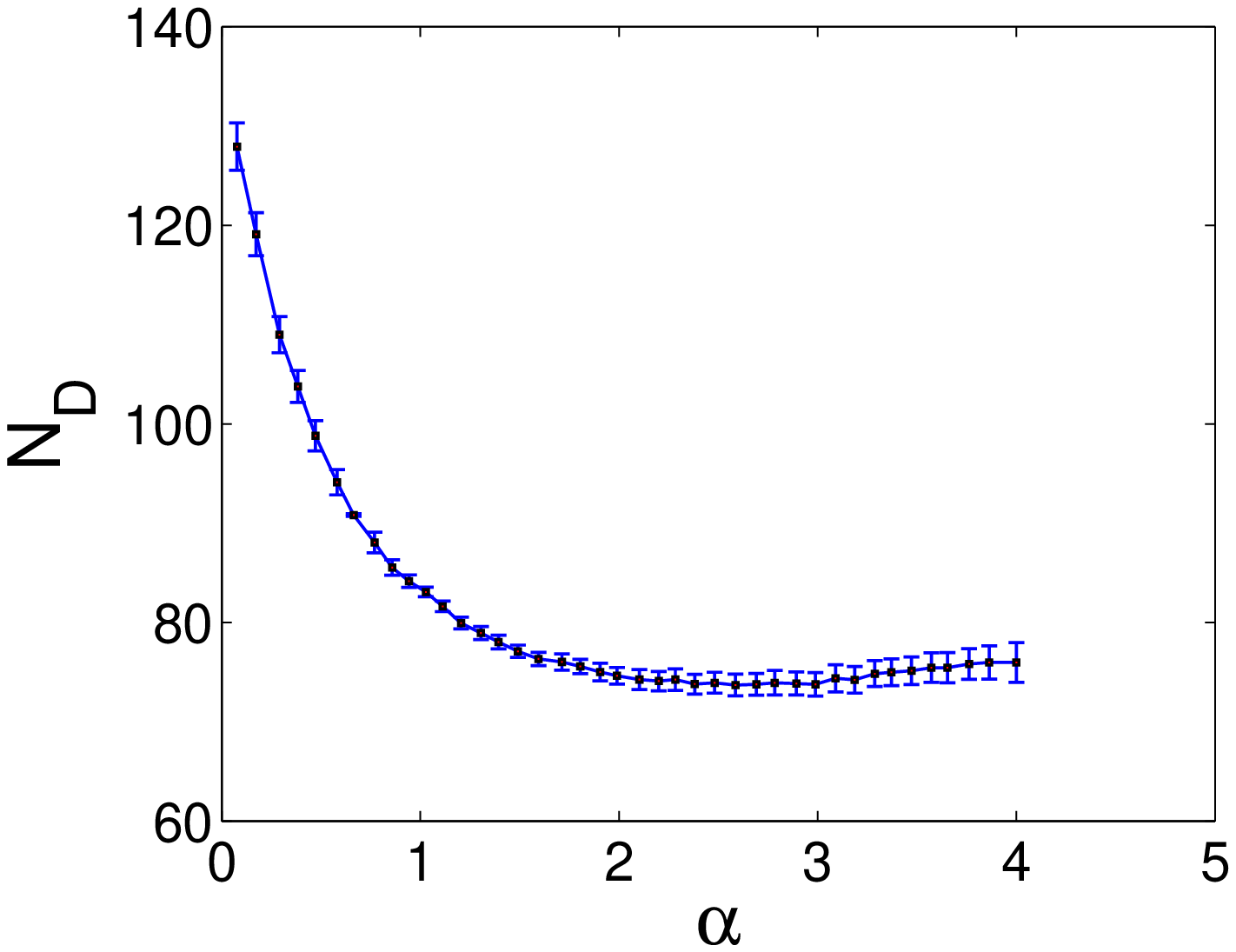}
\includegraphics[width=6.1cm]{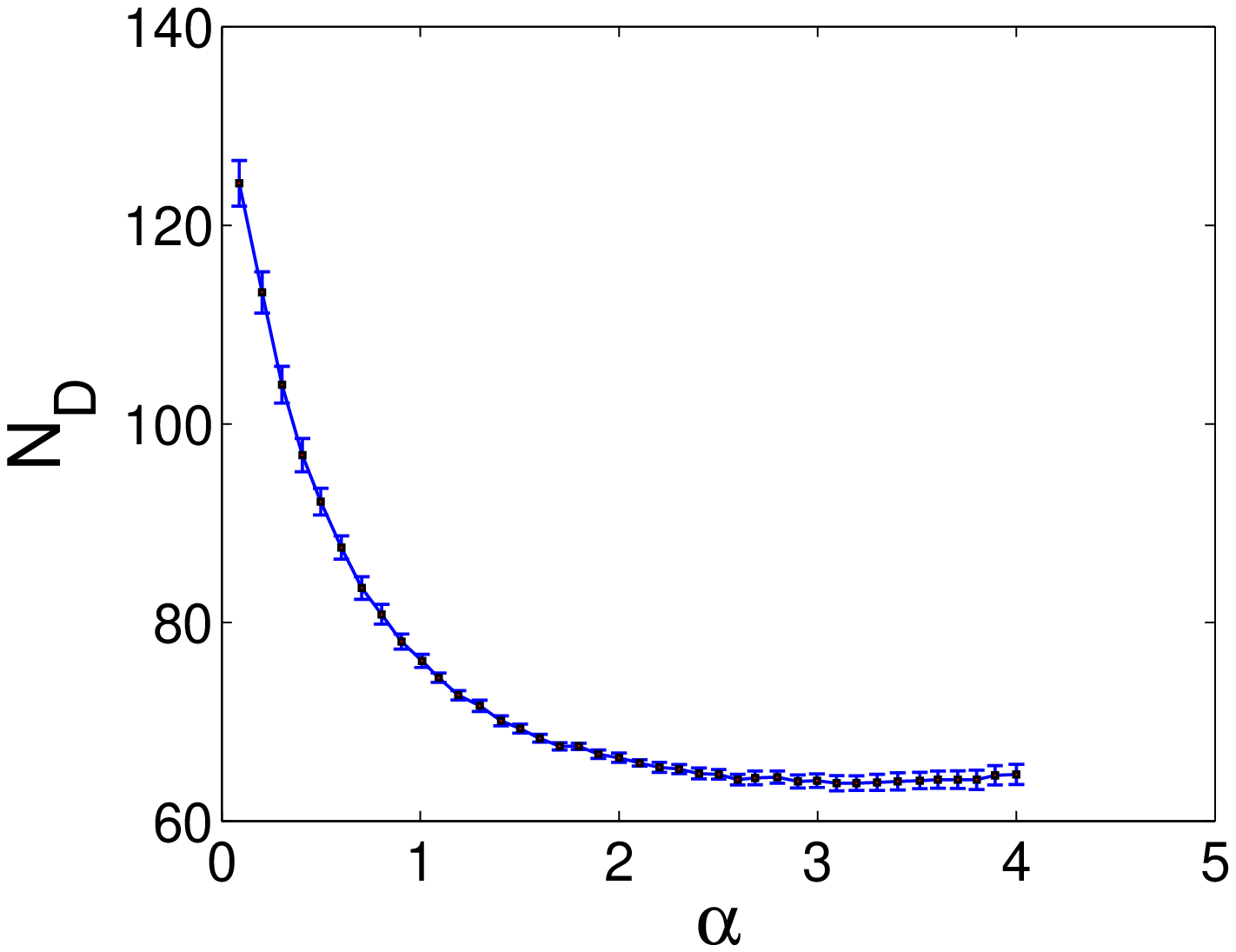}
\caption{(Color online) Average number of defects for fixed transition time $T$ as a function of the crossing exponent $\alpha$: (a) T=20; (b) T=40; (c) T=60; (d) T=80. The error bars are taken as the standard deviations of each set of data at fixed $T$ and $\alpha$. The solid lines connect the points and are only a guide to the eye.}
\label{fig:meanT}
\end{center}
\end{figure*}

We also performed numerical calculations for non-linear quenches.  For $\alpha=4$, the expected exponent is $\chi=2/5$. This is confirmed by the numerical results shown in the bottom panel of Fig.~\ref{fig:quench_lin_sq}.
After fitting the numerical data with the prediction given in Eq.~\eqref{eq:nd} we obtain the estimate $\chi_{fit} =0.408\pm 0.008$ in perfect agreement with the predicted result.

\subsection{Optimal exponent $\alpha$}
\label{sec:optalpha}
In this section we discuss the question: is it possible to find an optimal exponent $\alpha$ such that we minimise the number of defects produced with the constraint that the total traversing time $T$ is kept fixed?
The answer to this question was first given by Barankov and Polkovnikov \cite{Barankov2008}. They showed that the optimal exponent $\alpha_{opt}$ scales with universal critical exponents:
 \begin{equation}
 \label{eq:alphaopt}
\alpha_{opt}\approx - \frac{1}{\mu}\ln\left[\frac{1}{C T}\ln(CT)\right]
\end{equation}
where $C$ is a non-universal constant.
The corresponding scaling of the number of defects is greatly reduced with respect to the original one \eqref{eq:nd}:
\begin{equation}
 \label{eq:nopt}
N_{D,opt}\sim\left[\frac{1}{CT}\ln(CT)\right]^{\nu/\mu}.
\end{equation}
 The results shown in Eqs.~\eqref{eq:alphaopt} and \eqref{eq:nopt} are quite remarkable:  they show that the optimal passage exponent follows simple scaling relations related to universality. In the rest of this section we will compare predictions~\eqref{eq:alphaopt} and \eqref{eq:nopt} with our numerical simulations, and in the following we will show that by employing optimal control techniques we can reduce the number of defects even further.

In order to compare our numerical simulations with the predictions of Ref.~\cite{Barankov2008}, we modified the quench function $\eps(t)$ of Eq.~\eqref{eq:quench1} so that the total quenching time $T$ for going from 
$\eps(t_{in})=-2$ to $\eps(t_{in})=5$ is fixed a priori. In this setting, the function Eq.~\eqref{eq:quench1} is still valid, but the corresponding quench rate $\tau_Q$ now depends on $T$ and $\alpha$:
\begin{equation}
\tau_Q= \frac{T}{2^{1/\alpha}+5^{1/\alpha}}.
\end{equation}
The expressions for $t_{in}$ and $t_{fin}$ of Eq.~\eqref{eq:tintfin} remain unchanged.

\begin{figure}[b]
\begin{center}
\includegraphics[width=0.8 \columnwidth ]{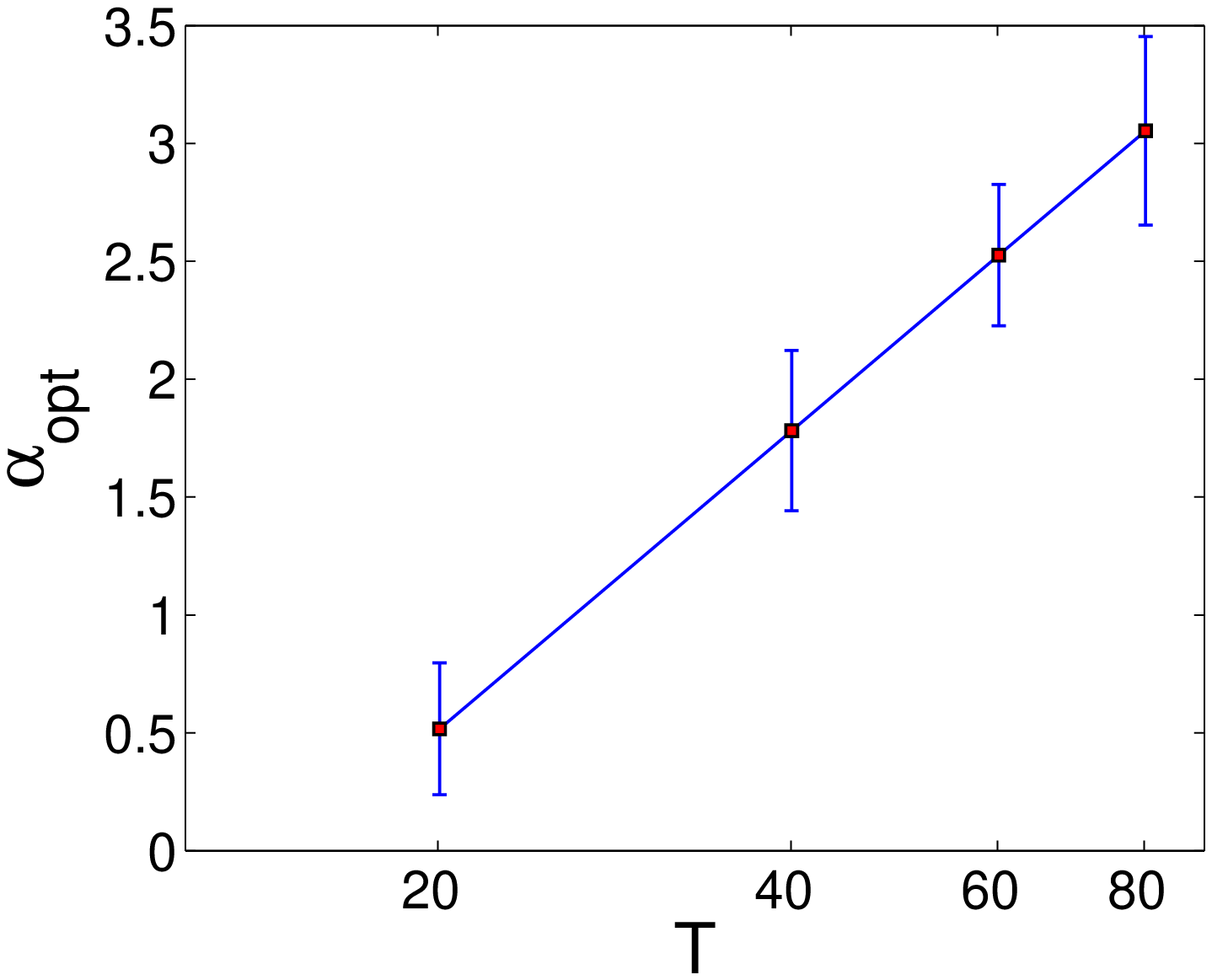}
~~~
\includegraphics[width=0.8 \columnwidth ]{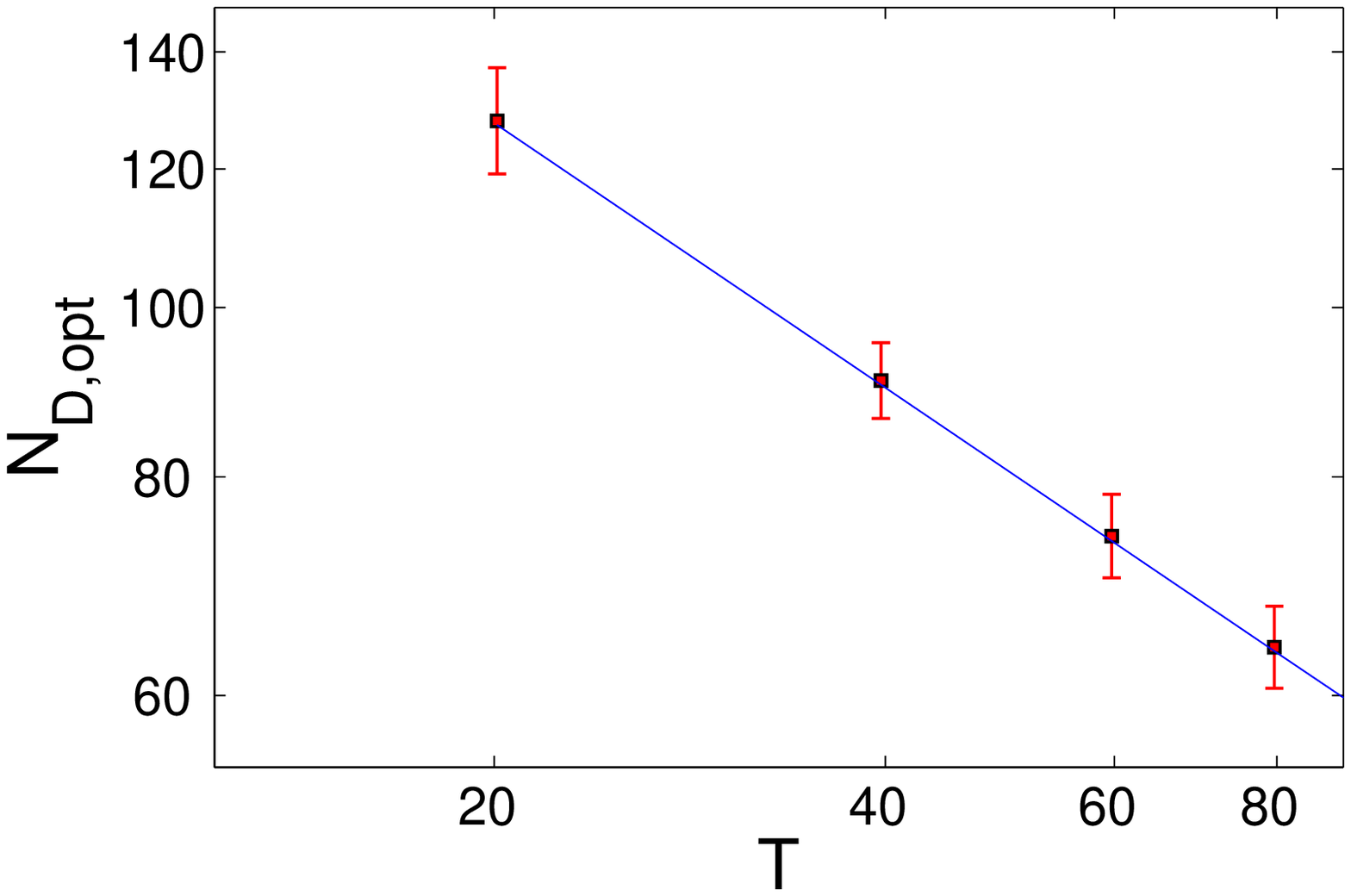}
\caption{(Color online) Top panel: Optimal traversing exponents $\alpha_{opt}$ (symbols) found from the minimisation of the data in Fig.~\ref{fig:meanT} versus the total time $T$ in semi-logarithmic scale. The solid line represents the best-fitting function $\tilde\alpha_{opt}$. Bottom panel: average number of defects $N_{D,opt}$ (symbols) as a function of the total time $T$ in log-log scale. The solid line is the best-fitting function $\tilde N_{D,opt}$ (see Eq.~\eqref{eq:Ndopt}).}
\label{fig:opt}
\end{center}
\end{figure}

For $T=20,40,60,80$ we vary $\alpha$ and compute the average number of defects $N_D$. The results are shown in Fig.~\ref{fig:meanT}.
For small values of $T$ we observe a clear optimal value $\alpha$ where the number of defects $N_D$ are minimised. As $T$ increases the minimum is very shallow and for $T>80$ we do not observe any clear minimum and $N_D$ decays to an asymptotic value.
To find the optimal values $\alpha_{opt}$ and $N_{D,opt}$ we interpolate the data with cubic splines. The estimates thus obtained are illustrated in Fig.~\ref{fig:opt}. In the top panel we show the estimated $\alpha_{opt}$ as a function of $T$ in a semi-logarithmic scale. The data points show a clear linear behaviour, thus we fit them with a simplified fitting function:
\begin{equation}
\tilde\alpha_{opt} =  A \ln\left[C T\right].
\end{equation}
We are therefore assuming that for a limited range of time lapses $T$, the double logarithmic term in Eq.~\eqref{eq:alphaopt} can be neglected.
After fitting the data, we extract the estimate for the prefactor: $A\simeq1.8 \pm 0.1$. This is quite in disagreement with the expected result $1/\mu=1$. The full model of Eq.~\eqref{eq:alphaopt} would not give a straight line in this scale and in fact does not agree with our numerical simulations.
In the small range of values of $T$ we were able to analyse, $\alpha_{opt}$ is well described by a power law of the total time $T$. We believe that the full model of Eq.~\eqref{eq:alphaopt} would be more appropriate for larger values of $T$.
However, in our numerical simulations, as we show in Fig.~\ref{fig:meanT}, we cannot take larger values of $T$ as it is impossible for us to accurately identify a minimum.

\subsection{Optimal number of defects $N_{D,opt}$}

We now turn to the analysis of the optimised average number of defects. The results of the numerical calculations are shown in the bottom panel of Fig.~\ref{fig:opt}. As we would like to compare the numerical data with the prediction of Eq.~\eqref{eq:nopt}, we plot the data in log-log scale. As before, we observe that the data show a clear linear scaling and therefore we fit them with a simple power law:
\begin{equation}
\label{eq:Ndopt}
\tilde N_{D,opt}=\left[\frac{1}{CT}\right]^\zeta.
\end{equation}
The fitting gives the estimate $\zeta \simeq 0.503 \pm 0.005$ which is in strong agreement with the prediction $\nu/\mu=0.5$ from Ref.~\cite{Barankov2008}. Therefore our numerical data for the average number of defects is well described by theoretical scaling relations. In contrast to the data for $\alpha_{opt}$, we find that  $ N_{D,opt}$ is less sensitive to the limited range of $T$.

\section{Optimised quenches}
\label{sec:optimal}
In this section we want to find strategies to minimise the production of defects, for a fixed  time $T$ for crossing the phase transition, by tailoring the time dependence of the reduced temperature $\eps(t)$. We go beyond the simple power law dependence presented in Eq.~\eqref{eq:quench1} and add a correction term $f(t)$ to it\footnote{The form in Eq.~\eqref{eq:quench2} implies that $\eps(0)=0$ regardless of the optimising function. We have also tried the following ansatz that relaxes the previous constraint: $\eps(t) = \frac{t}{|t|} \left(\frac{|t|}{\tau_Q} \right)^\alpha+f(t)$ which however gives poorer results.}:
\begin{equation}
\label{eq:quench2}
\eps(t) = \frac{t}{|t|} \left(\frac{|t|}{\tau_Q} \right)^\alpha[1+f(t)].
\end{equation}
We require that $|f(t_{in})|, |f(t_{fin})|\ll 1$ so that the initial and final values of the reduced temperature $\eps$ coincide approximately with the previously used values. Our task is then to find the function $f(t)$ that reduces the average number of defects created. This is a typical problem of optimal control (see for example \cite{krotov}) that has been recently employed for efficient cooling of many-body systems~\cite{chamon}. 
There are many algorithms that can be employed for this task and that could in principle guarantee monotonic decrease of the target cost function, in this case the average number of defects. We, however, use a simple yet powerful procedure inspired by the CRAB algorithm that was designed originally for optimising the dynamics of many-body quantum systems \cite{CRAB}.
The basic idea is to decompose the correction $f(t)$ as a linear superposition of trigonometric functions:
\begin{equation}
f(t)=\frac{1}{\lambda(t)}\sum_{n=1}^{n_{max}} A_n \cos\omega_n t+B_n \sin\omega_n t,
\end{equation}
where $n_{max}$ is the total number of frequencies $\omega_n$ that generate the correction $f(t)$; $A_n$ and $B_n$ are the amplitudes of the oscillating terms and we impose the following constraints:
\begin{equation}
\label{eq:f}
A_n^2\le 1;\quad B_n^2\le 1,
\end{equation}
which ensures that the optimisation algorithm will not yield oscillating functions with large amplitudes; finally the function $\lambda(t)$ forces the correction function to be smooth at the boundaries $t_{in}$ and $t_{fin}$. Although the specific form of $\lambda(t)$ is not crucial for the optimisation, we use the function:
\begin{equation}
\lambda(t)=1+ \Lambda\left[ e^{-(t-t_{in})^2} +e^{-(t-t_{fin})^2}\right]
\end{equation}
with the parameter $\Lambda=100$ forcing the control function $f(t)$ to be very small at the two endpoints.
For the frequencies appearing in Eq.~\eqref{eq:f}, we choose:
\begin{equation}
\omega_n = \frac{2\pi n}{T}.
\end{equation}

\begin{figure}[b]
\begin{center}
\includegraphics[width=0.9 \columnwidth ]{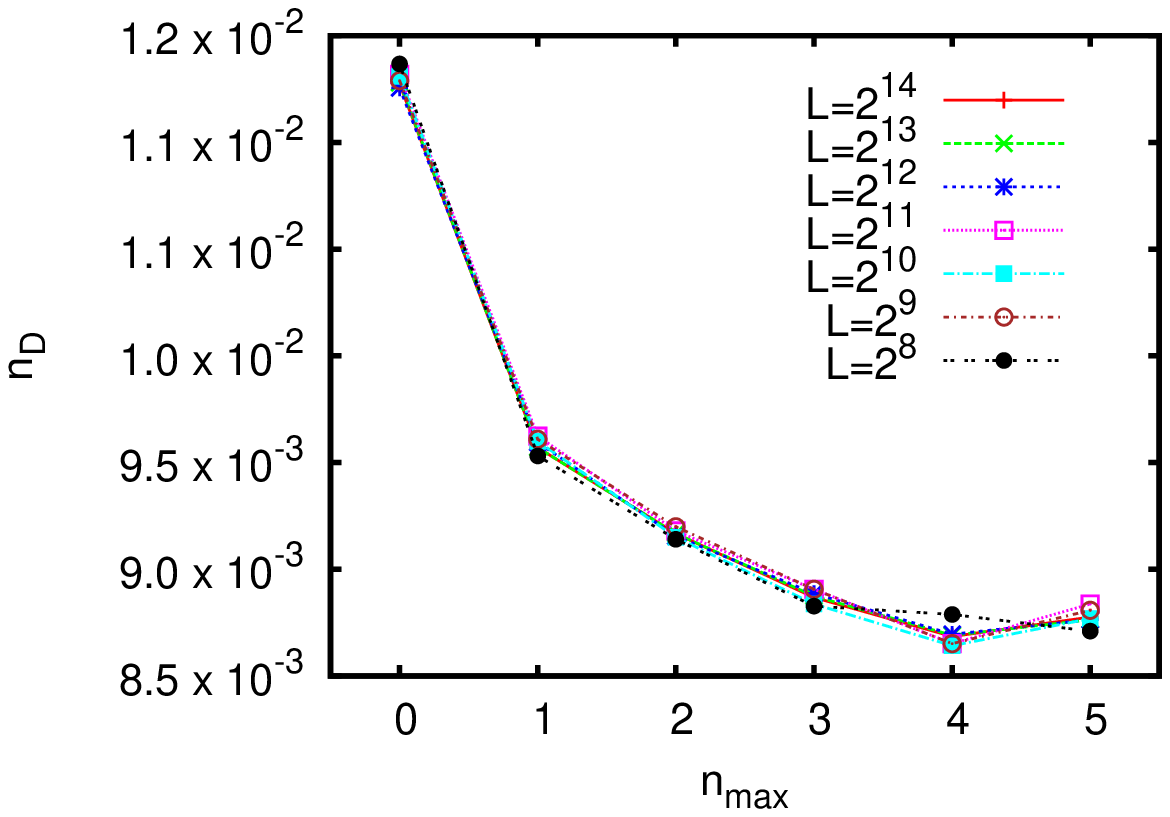}
~~~
\includegraphics[width=0.9 \columnwidth ]{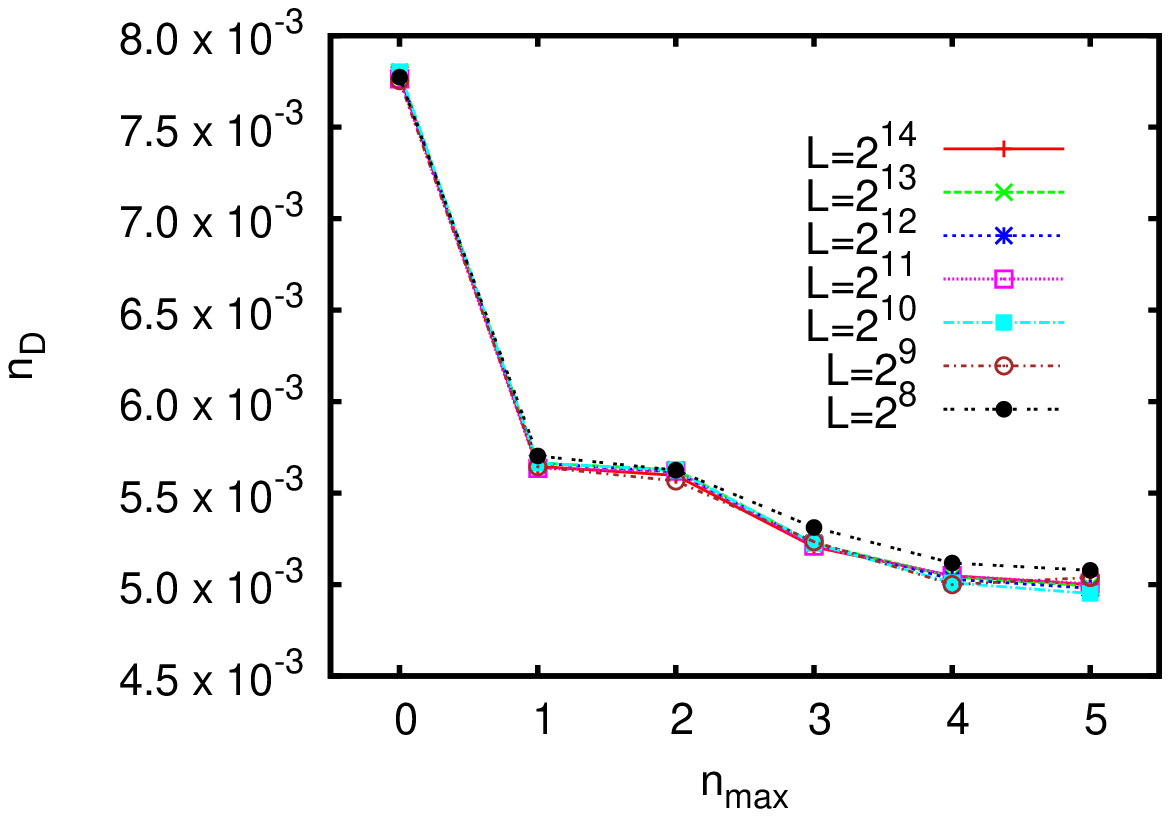}
\caption{(Color online) Optimised density of defects for $T=10$ (top panel) and $T=20$ (bottom panel) versus the number of frequencies $n_{max}$ in the optimisation algorithm. The different symbols are results obtained for different systems sizes $N$ ranging from $2^8$ to $2^{14}$. Straight lines connecting symbols are only a guide to the eye.}
\label{fig:OC}
\end{center}
\end{figure}

We first considered $T=20$ for concreteness. From the analysis in Sec.~\ref{sec:optalpha}, we know that the best exponent for the non-linear quench for $T=20$ is $\alpha=0.6$. This setting gives an average number of defects of $N_D\simeq 128\pm 1$.  We used standard Matlab minimisation routines to find the best values $A_n$ and $B_n$. The results are summarised in Fig.~\ref{fig:OC}.

\begin{figure}[h]
\begin{center}
\includegraphics[width=0.99 \columnwidth ]{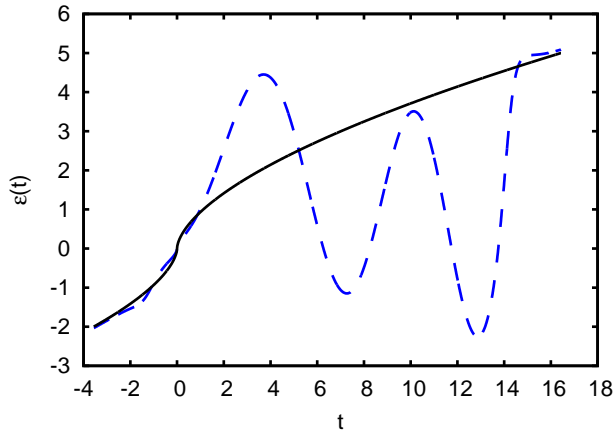}
\caption{(Color online) Optimal control results. Comparison of the optimised pulses $\eps(t)$ for $n_{max}=4$ (dashed line) with the original power-law dependence (solid line). We set $T=20$ and $\alpha=0.6$.}
\label{fig:opteps}
\end{center}
\end{figure}

%
The best result is for $n_{max}=5$, corresponding to 10 optimisation parameters, yielding $N_D=81.9\pm 0.2$ that is more than 40\% less then the non-optimised result.
For larger $n_{max}$ we cannot find better results as the number of free parameters is too large for the optimisation routines.

The resulting optimised time-dependence of the reduced temperature $\eps(t)$ is shown in Fig.~\ref{fig:opteps} and compared with the function $\eps(t)$ without optimisation. Similar to other optimal control results~\cite{DeChiara2008}, the control function exhibits non adiabatic oscillations that eventually lead to a reduction of the number of defects. While in the quantum scenario, as for example analysed in Ref.~\cite{DeChiara2008}, this is ascribed to constructive interference of many paths leading to the desired target state, in our classical model this might be interpreted as constructive interference of classical waves reducing the number of defects created. 
It is interesting to notice that the reduced temperature does not change monotonically and actually oscillates around zero a number of times. In terms of the Landau potential Eq.~\eqref{eq:landau}, the system evolves back and forth from a potential with a single minimum at zero order parameter $\varphi=0, \eps<0$ to a potential with two minima $\varphi\neq 0, \eps>0$.

We have extended our analysis to different system sizes $N$ ranging from $2^8$ to $2^{14}$ and also to a different total time $T=10$. The latter results have been obtained optimising the non-linear quench \eqref{eq:quench2} with $\alpha=0.5$. The results for the density of defects $n_D=N_D/N$ are shown in Fig.~\ref{fig:OC}. The data reveal that the optimised pulses are not very sensitive to the size of the system. Therefore, the performance of our optimisation protocol does not depend strongly on the exact number of particles in the system.

Our optimisation protocol is also quite robust to small imperfections in the coefficients $A$ and $B$. After perturbing these coefficients by random time-independent fluctuations of magnitude smaller than 1\% we find, on average, an increase in the number of defects by 3\%. 

\section{Conclusions}
\label{sec:conclusions}
In summary, we have provided numerical evidence that the total number of defects created during the crossing of a second order phase transition can be effectively reduced by appropriately tailoring the time dependence of the reduced temperature $\eps(t)$. This optimisation is only valid for a finite system, which is the relevant case for experiments. In the thermodynamical limit, the results presented in Ref.~\cite{Barankov2008} should remain valid: the optimal time dependence in the vicinity of the critical point should be a power law with an exponent $\alpha$ fulfilling universal scaling relations.

Our optimisation takes place in a open-system scenario, in which the system is always in contact with a thermal reservoir. In our simulations, this is embodied by the Langevin forces and the friction term. It is thus remarkable that a simple and intuitive technique as CRAB works in this non-ideal case. Moreover, as the number of frequencies is kept small, the bandwidth of the control function $f(t)$ can be kept under control for a realistic implementation.

{
Finally, we would like to stress that our work could be applied in experiments with classical systems undergoing 1D structural phase transitions of the second order such as those occurring for cold ions in highly anisotropic traps~\cite{exp_ions}.
}

\begin{acknowledgements}
We would like to thank T. Calarco, S. Montangero, and G. Morigi for useful
discussions, and C. Di Franco, A. Polkovnikov and A. Xuereb for their critical reading of the manuscript. We acknowledge the John Templeton Foundation (grant ID 43467) and EPSRC for financial support.
\end{acknowledgements}


\end{document}